\documentclass{llncs}

\usepackage{url}
\usepackage{breakurl} 
\usepackage[breaklinks]{hyperref}
\usepackage[table]{xcolor} 
\newcolumntype{?}{!{\vrule width 0.75pt}}
\usepackage{makecell}
\usepackage[T1]{fontenc} 
\usepackage[utf8]{inputenc}
\usepackage{times}
\usepackage{xstring}
\usepackage{textcomp}

\usepackage{units}
\usepackage{bytefield}
\usepackage{balance}
\usepackage[inline]{enumitem}
\makeatletter
\renewcommand\large{\@setfontsize\large\@xiipt{15\p@}}
\makeatother
\usepackage[tracking=true]{microtype}
\usepackage{booktabs}
\usepackage{tikz}
\usetikzlibrary{arrows,positioning}
\usepackage{cite}
\usepackage{multirow}
\usepackage{rotating}
\usepackage{subcaption}
\usepackage{tikz-qtree}
\usetikzlibrary{shadows,trees}

\tikzset{font=\tiny,
    edge from parent fork down,
    level distance=1.2cm,
    every node/.style={
        rectangle,rounded corners,
        minimum height=4mm,
        draw=black,
        very thick,
        align=left,
        text depth = 0pt
    },
    edge from parent/.style={
        draw=black,
        thick
    }
}

\newcommand{\ie}{i.e.,}
\newcommand{\eg}{e.g.,}
\newcommand{\etal}{et~al\@ifnextchar.{}{.\@}}
\newcommand{\etc}{etc\@ifnextchar.{}{.\@}}

\newcommand{\sref}[1]{Section~\ref{#1}} 
\newcommand{\fig}[1]{Figure~\ref{#1}}
\newcommand{\tab}[1]{Table~\ref{#1}} 

\newcommand{\afblock}[1]{\noindent{\textbf{#1}}}
\newcommand{\takeaway}[1]{\noindent{\textbf{Takeaway.}} \textit{#1}}

\usepackage{makeidx}  %
\newcommand{\icmpNetDetails}[1]{%
\IfStrEqCase{#1}{%
{type:0-UASs}{58}%
{type:0-UIPs}{301}%
{type:0-code:0-UASs}{45}%
{type:0-code:0-UIPs}{140}%
{type:0-code:161-UASs}{1}%
{type:0-code:161-UIPs}{1}%
{type:0-code:3-UASs}{4}%
{type:0-code:3-UIPs}{88}%
{type:0-code:80-UASs}{10}%
{type:0-code:80-UIPs}{72}%
{type:1-code:187-UASs}{31}%
{type:1-code:187-UIPs}{559}%
{type:104-code:58-UASs}{1}%
{type:104-code:58-UIPs}{1}%
{type:11-UASs}{18.40K}%
{type:11-UIPs}{455.13K}%
{type:11-code:0-UASs}{18.40K}%
{type:11-code:0-UIPs}{447.95K}%
{type:11-code:1-UASs}{30}%
{type:11-code:1-UIPs}{7.19K}%
{type:11-code:128-UASs}{1}%
{type:11-code:128-UIPs}{2}%
{type:118-code:245-UASs}{1}%
{type:118-code:245-UIPs}{1}%
{type:12-UASs}{9}%
{type:12-UIPs}{16}%
{type:12-code:0-UASs}{9}%
{type:12-code:0-UIPs}{16}%
{type:120-code:1-UASs}{1}%
{type:120-code:1-UIPs}{1}%
{type:128-code:1-UASs}{1}%
{type:128-code:1-UIPs}{1}%
{type:13-UASs}{6}%
{type:13-UIPs}{9}%
{type:13-code:0-UASs}{6}%
{type:13-code:0-UIPs}{9}%
{type:137-code:226-UASs}{1}%
{type:137-code:226-UIPs}{6}%
{type:149-code:87-UASs}{1}%
{type:149-code:87-UIPs}{1}%
{type:154-code:219-UASs}{1}%
{type:154-code:219-UIPs}{1}%
{type:164-code:140-UASs}{1}%
{type:164-code:140-UIPs}{1}%
{type:168-code:200-UASs}{1}%
{type:168-code:200-UIPs}{1}%
{type:17-UASs}{1}%
{type:17-UIPs}{1}%
{type:17-code:0-UASs}{1}%
{type:17-code:0-UIPs}{1}%
{type:174-code:188-UASs}{1}%
{type:174-code:188-UIPs}{1}%
{type:176-code:175-UASs}{1}%
{type:176-code:175-UIPs}{1}%
{type:187-code:9-UASs}{1}%
{type:187-code:9-UIPs}{1}%
{type:188-code:137-UASs}{1}%
{type:188-code:137-UIPs}{1}%
{type:189-code:253-UASs}{1}%
{type:189-code:253-UIPs}{1}%
{type:191-code:13-UASs}{1}%
{type:191-code:13-UIPs}{1}%
{type:2-code:3-UASs}{1}%
{type:2-code:3-UIPs}{2}%
{type:201-code:164-UASs}{1}%
{type:201-code:164-UIPs}{1}%
{type:212-code:1-UASs}{1}%
{type:212-code:1-UIPs}{1}%
{type:212-code:36-UASs}{1}%
{type:212-code:36-UIPs}{1}%
{type:227-code:118-UASs}{1}%
{type:227-code:118-UIPs}{1}%
{type:227-code:22-UASs}{1}%
{type:227-code:22-UIPs}{1}%
{type:230-code:253-UASs}{1}%
{type:230-code:253-UIPs}{1}%
{type:236-code:3-UASs}{1}%
{type:236-code:3-UIPs}{1}%
{type:236-code:61-UASs}{1}%
{type:236-code:61-UIPs}{1}%
{type:239-code:80-UASs}{1}%
{type:239-code:80-UIPs}{1}%
{type:241-code:211-UASs}{1}%
{type:241-code:211-UIPs}{1}%
{type:241-code:28-UASs}{1}%
{type:241-code:28-UIPs}{1}%
{type:245-code:105-UASs}{1}%
{type:245-code:105-UIPs}{1}%
{type:248-code:48-UASs}{1}%
{type:248-code:48-UIPs}{1}%
{type:25-code:119-UASs}{1}%
{type:25-code:119-UIPs}{1}%
{type:255-code:255-UASs}{2}%
{type:255-code:255-UIPs}{9}%
{type:3-UASs}{52.92K}%
{type:3-UIPs}{170.30M}%
{type:3-code:0-UASs}{5.67K}%
{type:3-code:0-UIPs}{66.76K}%
{type:3-code:1-UASs}{32.57K}%
{type:3-code:1-UIPs}{7.29M}%
{type:3-code:10-UASs}{20.24K}%
{type:3-code:10-UIPs}{4.28M}%
{type:3-code:109-UASs}{1}%
{type:3-code:109-UIPs}{2}%
{type:3-code:11-UASs}{6}%
{type:3-code:11-UIPs}{7}%
{type:3-code:13-UASs}{17.55K}%
{type:3-code:13-UIPs}{21.26M}%
{type:3-code:168-UASs}{1}%
{type:3-code:168-UIPs}{1}%
{type:3-code:2-UASs}{925}%
{type:3-code:2-UIPs}{25.19K}%
{type:3-code:211-UASs}{1}%
{type:3-code:211-UIPs}{1}%
{type:3-code:214-UASs}{1}%
{type:3-code:214-UIPs}{1}%
{type:3-code:3-UASs}{49.91K}%
{type:3-code:3-UIPs}{139.12M}%
{type:3-code:4-UASs}{332}%
{type:3-code:4-UIPs}{1.03K}%
{type:3-code:6-UASs}{2}%
{type:3-code:6-UIPs}{3}%
{type:3-code:7-UASs}{13}%
{type:3-code:7-UIPs}{29}%
{type:3-code:8-UASs}{1}%
{type:3-code:8-UIPs}{1}%
{type:3-code:9-UASs}{288}%
{type:3-code:9-UIPs}{1.20K}%
{type:3-code:95-UASs}{1}%
{type:3-code:95-UIPs}{1}%
{type:32-code:116-UASs}{1}%
{type:32-code:116-UIPs}{1}%
{type:35-code:3-UASs}{1}%
{type:35-code:3-UIPs}{1}%
{type:4-UASs}{364}%
{type:4-UIPs}{2.65K}%
{type:4-code:0-UASs}{364}%
{type:4-code:0-UIPs}{2.65K}%
{type:46-code:48-UASs}{1}%
{type:46-code:48-UIPs}{1}%
{type:5-UASs}{2.29K}%
{type:5-UIPs}{243.25K}%
{type:5-code:0-UASs}{238}%
{type:5-code:0-UIPs}{490}%
{type:5-code:1-UASs}{2.20K}%
{type:5-code:1-UIPs}{242.78K}%
{type:50-code:46-UASs}{1}%
{type:50-code:46-UIPs}{4}%
{type:69-code:0-UASs}{1}%
{type:69-code:0-UIPs}{1}%
{type:74-code:58-UASs}{1}%
{type:74-code:58-UIPs}{1}%
{type:8-UASs}{861}%
{type:8-UIPs}{10.64K}%
{type:8-code:0-UASs}{840}%
{type:8-code:0-UIPs}{10.57K}%
{type:8-code:123-UASs}{1}%
{type:8-code:123-UIPs}{1}%
{type:8-code:19-UASs}{35}%
{type:8-code:19-UIPs}{66}%
{type:8-code:3-UASs}{2}%
{type:8-code:3-UIPs}{2}%
{type:8-code:82-UASs}{1}%
{type:8-code:82-UIPs}{1}%
{type:8-code:9-UASs}{4}%
{type:8-code:9-UIPs}{4}%
{type:91-code:144-UASs}{1}%
{type:91-code:144-UIPs}{1}%
{type:Other-UASs}{43}%
{type:Other-UIPs}{606}%
}%
[\textcolor{red}{#1 not found in icmpNetDetails}]%
}

\newcommand{\icmp}[1]{%
\IfStrEqCase{#1}{%
{num}{75}%
{total}{637.50M}%
{type:0-code:0-total}{198}%
{type:0-code:161-total}{2}%
{type:0-code:3-total}{5.75K}%
{type:0-code:80-total}{137}%
{type:0-total}{6.08K}%
{type:11-code:0-total}{139.52M}%
{type:11-code:1-total}{7.31K}%
{type:11-code:128-total}{3}%
{type:11-total}{139.53M}%
{type:12-code:0-total}{20}%
{type:12-total}{20}%
{type:13-code:0-total}{73}%
{type:13-total}{73}%
{type:17-code:0-total}{4}%
{type:17-total}{4}%
{type:3-code:0-total}{17.94M}%
{type:3-code:1-total}{107.15M}%
{type:3-code:10-total}{23.07M}%
{type:3-code:109-total}{2}%
{type:3-code:11-total}{25}%
{type:3-code:13-total}{71.70M}%
{type:3-code:168-total}{1}%
{type:3-code:2-total}{51.04K}%
{type:3-code:211-total}{1}%
{type:3-code:214-total}{1}%
{type:3-code:3-total}{256.72M}%
{type:3-code:4-total}{26.66K}%
{type:3-code:6-total}{6}%
{type:3-code:7-total}{336}%
{type:3-code:8-total}{2}%
{type:3-code:9-total}{26.28K}%
{type:3-code:95-total}{1}%
{type:3-total}{476.68M}%
{type:4-code:0-total}{46.18K}%
{type:4-total}{46.18K}%
{type:5-code:0-total}{105.78K}%
{type:5-code:1-total}{18.01M}%
{type:5-total}{18.12M}%
{type:8-code:0-total}{3.12M}%
{type:8-code:123-total}{1}%
{type:8-code:19-total}{96}%
{type:8-code:3-total}{4}%
{type:8-code:82-total}{3}%
{type:8-code:9-total}{739}%
{type:8-total}{3.12M}%
{type:Other-total}{1.48K}%
{type:Other_code:Other-total}{1.48K}%
}%
[\textcolor{red}{#1 not found in icmp}]%
}

\newcommand{\zmaptcp}[1]{%
\IfStrEqCase{#1}{%
{num}{23}%
{total}{98.33M}%
{type:0-code:0-total}{3}%
{type:0-code:80-total}{131}%
{type:0-total}{134}%
{type:11-code:0-total}{37.30M}%
{type:11-code:1-total}{214}%
{type:11-total}{37.30M}%
{type:12-code:0-total}{8}%
{type:12-total}{8}%
{type:13-code:0-total}{20}%
{type:13-total}{20}%
{type:3-code:0-total}{4.20M}%
{type:3-code:1-total}{25.49M}%
{type:3-code:10-total}{1.36M}%
{type:3-code:11-total}{7}%
{type:3-code:13-total}{18.86M}%
{type:3-code:2-total}{16.37K}%
{type:3-code:3-total}{5.50M}%
{type:3-code:6-total}{3}%
{type:3-code:7-total}{66}%
{type:3-code:9-total}{5.28K}%
{type:3-total}{55.44M}%
{type:4-code:0-total}{3.81K}%
{type:4-total}{3.81K}%
{type:5-code:0-total}{26.26K}%
{type:5-code:1-total}{5.54M}%
{type:5-total}{5.57M}%
{type:8-code:0-total}{24.90K}%
{type:8-code:19-total}{92}%
{type:8-code:9-total}{174}%
{type:8-total}{25.17K}%
{type:Other-total}{1}%
{type:Other_code:Other-total}{1}%
}%
[\textcolor{red}{#1 not found in zmaptcp}]%
}

\newcommand{\zmaptcpdetail}[1]{%
\IfStrEqCase{#1}{%
{type:0-UASs}{10}%
{type:0-UIPs}{74}%
{type:0-code:0-UASs}{1}%
{type:0-code:0-UIPs}{3}%
{type:0-code:80-UASs}{9}%
{type:0-code:80-UIPs}{71}%
{type:11-UASs}{16.65K}%
{type:11-UIPs}{338.06K}%
{type:11-code:0-UASs}{16.65K}%
{type:11-code:0-UIPs}{337.86K}%
{type:11-code:1-UASs}{10}%
{type:11-code:1-UIPs}{201}%
{type:12-UASs}{2}%
{type:12-UIPs}{8}%
{type:12-code:0-UASs}{2}%
{type:12-code:0-UIPs}{8}%
{type:13-UASs}{2}%
{type:13-UIPs}{2}%
{type:13-code:0-UASs}{2}%
{type:13-code:0-UIPs}{2}%
{type:187-code:9-UASs}{1}%
{type:187-code:9-UIPs}{1}%
{type:3-UASs}{36.60K}%
{type:3-UIPs}{24.96M}%
{type:3-code:0-UASs}{4.82K}%
{type:3-code:0-UIPs}{44.30K}%
{type:3-code:1-UASs}{29.84K}%
{type:3-code:1-UIPs}{4.57M}%
{type:3-code:10-UASs}{14.27K}%
{type:3-code:10-UIPs}{1.28M}%
{type:3-code:11-UASs}{6}%
{type:3-code:11-UIPs}{7}%
{type:3-code:13-UASs}{16.25K}%
{type:3-code:13-UIPs}{13.91M}%
{type:3-code:2-UASs}{606}%
{type:3-code:2-UIPs}{15.72K}%
{type:3-code:3-UASs}{9.19K}%
{type:3-code:3-UIPs}{5.15M}%
{type:3-code:6-UASs}{2}%
{type:3-code:6-UIPs}{3}%
{type:3-code:7-UASs}{8}%
{type:3-code:7-UIPs}{21}%
{type:3-code:9-UASs}{115}%
{type:3-code:9-UIPs}{236}%
{type:4-UASs}{168}%
{type:4-UIPs}{698}%
{type:4-code:0-UASs}{168}%
{type:4-code:0-UIPs}{698}%
{type:5-UASs}{1.89K}%
{type:5-UIPs}{28.01K}%
{type:5-code:0-UASs}{201}%
{type:5-code:0-UIPs}{399}%
{type:5-code:1-UASs}{1.81K}%
{type:5-code:1-UIPs}{27.62K}%
{type:8-UASs}{206}%
{type:8-UIPs}{631}%
{type:8-code:0-UASs}{177}%
{type:8-code:0-UIPs}{566}%
{type:8-code:19-UASs}{34}%
{type:8-code:19-UIPs}{65}%
{type:8-code:9-UASs}{1}%
{type:8-code:9-UIPs}{1}%
{type:Other-UASs}{1}%
{type:Other-UIPs}{1}%
}%
[\textcolor{red}{#1 not found in zmaptcpdetail}]%
}

\newcommand{\zmapquic}[1]{%
\IfStrEqCase{#1}{%
{num}{50}%
{total}{198.16M}%
{type:0-code:0-total}{74}%
{type:0-code:161-total}{1}%
{type:0-code:3-total}{2}%
{type:0-total}{77}%
{type:11-code:0-total}{29.00M}%
{type:11-code:1-total}{3.50K}%
{type:11-total}{29.01M}%
{type:12-code:0-total}{5}%
{type:12-total}{5}%
{type:13-code:0-total}{1}%
{type:13-total}{1}%
{type:3-code:0-total}{4.17M}%
{type:3-code:1-total}{24.45M}%
{type:3-code:10-total}{3.29M}%
{type:3-code:109-total}{1}%
{type:3-code:11-total}{6}%
{type:3-code:13-total}{16.50M}%
{type:3-code:2-total}{8.04K}%
{type:3-code:211-total}{1}%
{type:3-code:214-total}{1}%
{type:3-code:3-total}{117.56M}%
{type:3-code:4-total}{2.84K}%
{type:3-code:6-total}{1}%
{type:3-code:7-total}{74}%
{type:3-code:8-total}{1}%
{type:3-code:9-total}{7.47K}%
{type:3-total}{165.98M}%
{type:4-code:0-total}{1.80K}%
{type:4-total}{1.80K}%
{type:5-code:0-total}{25.09K}%
{type:5-code:1-total}{3.12M}%
{type:5-total}{3.15M}%
{type:8-code:0-total}{24.46K}%
{type:8-code:3-total}{2}%
{type:8-code:82-total}{1}%
{type:8-code:9-total}{52}%
{type:8-total}{24.52K}%
{type:Other-total}{599}%
{type:Other_code:Other-total}{599}%
}%
[\textcolor{red}{#1 not found in zmapquic}]%
}

\newcommand{\zmapquicdetail}[1]{%
\IfStrEqCase{#1}{%
{type:0-UASs}{37}%
{type:0-UIPs}{77}%
{type:0-code:0-UASs}{34}%
{type:0-code:0-UIPs}{74}%
{type:0-code:161-UASs}{1}%
{type:0-code:161-UIPs}{1}%
{type:0-code:3-UASs}{2}%
{type:0-code:3-UIPs}{2}%
{type:1-code:187-UASs}{28}%
{type:1-code:187-UIPs}{480}%
{type:11-UASs}{15.79K}%
{type:11-UIPs}{342.58K}%
{type:11-code:0-UASs}{15.79K}%
{type:11-code:0-UIPs}{339.07K}%
{type:11-code:1-UASs}{15}%
{type:11-code:1-UIPs}{3.50K}%
{type:118-code:245-UASs}{1}%
{type:118-code:245-UIPs}{1}%
{type:12-UASs}{5}%
{type:12-UIPs}{5}%
{type:12-code:0-UASs}{5}%
{type:12-code:0-UIPs}{5}%
{type:120-code:1-UASs}{1}%
{type:120-code:1-UIPs}{1}%
{type:128-code:1-UASs}{1}%
{type:128-code:1-UIPs}{1}%
{type:13-UASs}{1}%
{type:13-UIPs}{1}%
{type:13-code:0-UASs}{1}%
{type:13-code:0-UIPs}{1}%
{type:137-code:226-UASs}{1}%
{type:137-code:226-UIPs}{6}%
{type:154-code:219-UASs}{1}%
{type:154-code:219-UIPs}{1}%
{type:174-code:188-UASs}{1}%
{type:174-code:188-UIPs}{1}%
{type:188-code:137-UASs}{1}%
{type:188-code:137-UIPs}{1}%
{type:2-code:3-UASs}{1}%
{type:2-code:3-UIPs}{2}%
{type:201-code:164-UASs}{1}%
{type:201-code:164-UIPs}{1}%
{type:212-code:1-UASs}{1}%
{type:212-code:1-UIPs}{1}%
{type:227-code:118-UASs}{1}%
{type:227-code:118-UIPs}{1}%
{type:236-code:3-UASs}{1}%
{type:236-code:3-UIPs}{1}%
{type:245-code:105-UASs}{1}%
{type:245-code:105-UIPs}{1}%
{type:25-code:119-UASs}{1}%
{type:25-code:119-UIPs}{1}%
{type:255-code:255-UASs}{2}%
{type:255-code:255-UIPs}{5}%
{type:3-UASs}{50.85K}%
{type:3-UIPs}{134.72M}%
{type:3-code:0-UASs}{4.71K}%
{type:3-code:0-UIPs}{42.73K}%
{type:3-code:1-UASs}{25.53K}%
{type:3-code:1-UIPs}{3.78M}%
{type:3-code:10-UASs}{17.45K}%
{type:3-code:10-UIPs}{3.11M}%
{type:3-code:109-UASs}{1}%
{type:3-code:109-UIPs}{1}%
{type:3-code:11-UASs}{5}%
{type:3-code:11-UIPs}{6}%
{type:3-code:13-UASs}{16.50K}%
{type:3-code:13-UIPs}{14.57M}%
{type:3-code:2-UASs}{486}%
{type:3-code:2-UIPs}{7.55K}%
{type:3-code:211-UASs}{1}%
{type:3-code:211-UIPs}{1}%
{type:3-code:214-UASs}{1}%
{type:3-code:214-UIPs}{1}%
{type:3-code:3-UASs}{49.50K}%
{type:3-code:3-UIPs}{113.76M}%
{type:3-code:4-UASs}{260}%
{type:3-code:4-UIPs}{894}%
{type:3-code:6-UASs}{1}%
{type:3-code:6-UIPs}{1}%
{type:3-code:7-UASs}{12}%
{type:3-code:7-UIPs}{27}%
{type:3-code:8-UASs}{1}%
{type:3-code:8-UIPs}{1}%
{type:3-code:9-UASs}{232}%
{type:3-code:9-UIPs}{824}%
{type:32-code:116-UASs}{1}%
{type:32-code:116-UIPs}{1}%
{type:4-UASs}{140}%
{type:4-UIPs}{435}%
{type:4-code:0-UASs}{140}%
{type:4-code:0-UIPs}{435}%
{type:5-UASs}{1.61K}%
{type:5-UIPs}{24.66K}%
{type:5-code:0-UASs}{171}%
{type:5-code:0-UIPs}{354}%
{type:5-code:1-UASs}{1.54K}%
{type:5-code:1-UIPs}{24.31K}%
{type:50-code:46-UASs}{1}%
{type:50-code:46-UIPs}{1}%
{type:69-code:0-UASs}{1}%
{type:69-code:0-UIPs}{1}%
{type:74-code:58-UASs}{1}%
{type:74-code:58-UIPs}{1}%
{type:8-UASs}{101}%
{type:8-UIPs}{2.44K}%
{type:8-code:0-UASs}{99}%
{type:8-code:0-UIPs}{2.44K}%
{type:8-code:3-UASs}{2}%
{type:8-code:3-UIPs}{2}%
{type:8-code:82-UASs}{1}%
{type:8-code:82-UIPs}{1}%
{type:8-code:9-UASs}{2}%
{type:8-code:9-UIPs}{2}%
{type:91-code:144-UASs}{1}%
{type:91-code:144-UIPs}{1}%
{type:Other-UASs}{38}%
{type:Other-UIPs}{509}%
}%
[\textcolor{red}{#1 not found in zmapquicdetail}]%
}

\newcommand{\dnshtwo}[1]{%
\IfStrEqCase{#1}{%
{num}{17}%
{total}{5.28M}%
{type:11-code:0-total}{349.26K}%
{type:11-code:1-total}{1}%
{type:11-total}{349.26K}%
{type:3-code:0-total}{75.19K}%
{type:3-code:1-total}{483.00K}%
{type:3-code:10-total}{2.54M}%
{type:3-code:13-total}{40.39K}%
{type:3-code:2-total}{35}%
{type:3-code:3-total}{1.13M}%
{type:3-code:4-total}{13.82K}%
{type:3-code:7-total}{8}%
{type:3-code:9-total}{14}%
{type:3-total}{4.28M}%
{type:4-code:0-total}{59}%
{type:4-total}{59}%
{type:5-code:0-total}{16}%
{type:5-code:1-total}{2.99K}%
{type:5-total}{3.01K}%
{type:8-code:0-total}{647.02K}%
{type:8-code:9-total}{37}%
{type:8-total}{647.06K}%
{type:Other-total}{23}%
{type:Other_code:Other-total}{23}%
}%
[\textcolor{red}{#1 not found in dnshtwo}]%
}

\newcommand{\dnshtwodetail}[1]{%
\IfStrEqCase{#1}{%
{type:1-code:187-UASs}{3}%
{type:1-code:187-UIPs}{6}%
{type:11-UASs}{2.11K}%
{type:11-UIPs}{7.05K}%
{type:11-code:0-UASs}{2.11K}%
{type:11-code:0-UIPs}{7.05K}%
{type:11-code:1-UASs}{1}%
{type:11-code:1-UIPs}{1}%
{type:3-UASs}{9.39K}%
{type:3-UIPs}{402.77K}%
{type:3-code:0-UASs}{378}%
{type:3-code:0-UIPs}{866}%
{type:3-code:1-UASs}{6.17K}%
{type:3-code:1-UIPs}{34.20K}%
{type:3-code:10-UASs}{4.64K}%
{type:3-code:10-UIPs}{344.11K}%
{type:3-code:13-UASs}{1.14K}%
{type:3-code:13-UIPs}{7.18K}%
{type:3-code:2-UASs}{14}%
{type:3-code:2-UIPs}{20}%
{type:3-code:3-UASs}{1.76K}%
{type:3-code:3-UIPs}{16.49K}%
{type:3-code:4-UASs}{36}%
{type:3-code:4-UIPs}{55}%
{type:3-code:7-UASs}{1}%
{type:3-code:7-UIPs}{1}%
{type:3-code:9-UASs}{6}%
{type:3-code:9-UIPs}{7}%
{type:4-UASs}{10}%
{type:4-UIPs}{13}%
{type:4-code:0-UASs}{10}%
{type:4-code:0-UIPs}{13}%
{type:5-UASs}{68}%
{type:5-UIPs}{176}%
{type:5-code:0-UASs}{7}%
{type:5-code:0-UIPs}{7}%
{type:5-code:1-UASs}{61}%
{type:5-code:1-UIPs}{169}%
{type:8-UASs}{137}%
{type:8-UIPs}{450}%
{type:8-code:0-UASs}{137}%
{type:8-code:0-UIPs}{450}%
{type:8-code:9-UASs}{2}%
{type:8-code:9-UIPs}{2}%
{type:Other-UASs}{3}%
{type:Other-UIPs}{6}%
}%
[\textcolor{red}{#1 not found in dnshtwodetail}]%
}

\newcommand{\dnsquic}[1]{%
\IfStrEqCase{#1}{%
{num}{8}%
{total}{14.29K}%
{type:11-code:0-total}{7}%
{type:11-total}{7}%
{type:3-code:1-total}{1}%
{type:3-code:10-total}{358}%
{type:3-code:13-total}{1}%
{type:3-code:2-total}{3}%
{type:3-code:3-total}{8}%
{type:3-code:4-total}{1}%
{type:3-total}{372}%
{type:8-code:0-total}{13.91K}%
{type:8-total}{13.91K}%
{type:Other-total}{0}%
{type:Other_code:Other-total}{0}%
}%
[\textcolor{red}{#1 not found in dnsquic}]%
}

\newcommand{\dnsquicdetail}[1]{%
\IfStrEqCase{#1}{%
{type:11-UASs}{3}%
{type:11-UIPs}{5}%
{type:11-code:0-UASs}{3}%
{type:11-code:0-UIPs}{5}%
{type:3-UASs}{30}%
{type:3-UIPs}{42}%
{type:3-code:1-UASs}{1}%
{type:3-code:1-UIPs}{1}%
{type:3-code:10-UASs}{17}%
{type:3-code:10-UIPs}{29}%
{type:3-code:13-UASs}{1}%
{type:3-code:13-UIPs}{1}%
{type:3-code:2-UASs}{3}%
{type:3-code:2-UIPs}{3}%
{type:3-code:3-UASs}{7}%
{type:3-code:3-UIPs}{7}%
{type:3-code:4-UASs}{1}%
{type:3-code:4-UIPs}{1}%
{type:8-UASs}{73}%
{type:8-UIPs}{157}%
{type:8-code:0-UASs}{73}%
{type:8-code:0-UIPs}{157}%
}%
[\textcolor{red}{#1 not found in dnsquicdetail}]%
}

\newcommand{\icmpFULL}[1]{%
\IfStrEqCase{#1}{%
{num}{75}%
{total}{637.50M}%
{type:0-code:0-total}{198}%
{type:0-code:161-total}{2}%
{type:0-code:3-total}{5.75K}%
{type:0-code:80-total}{137}%
{type:1-code:187-total}{1.42K}%
{type:104-code:58-total}{1}%
{type:11-code:0-total}{139.52M}%
{type:11-code:1-total}{7.31K}%
{type:11-code:128-total}{3}%
{type:118-code:245-total}{1}%
{type:12-code:0-total}{20}%
{type:120-code:1-total}{1}%
{type:128-code:1-total}{1}%
{type:13-code:0-total}{73}%
{type:137-code:226-total}{12}%
{type:149-code:87-total}{1}%
{type:154-code:219-total}{1}%
{type:164-code:140-total}{1}%
{type:168-code:200-total}{1}%
{type:17-code:0-total}{4}%
{type:174-code:188-total}{1}%
{type:176-code:175-total}{1}%
{type:187-code:9-total}{1}%
{type:188-code:137-total}{1}%
{type:189-code:253-total}{1}%
{type:191-code:13-total}{1}%
{type:2-code:3-total}{2}%
{type:201-code:164-total}{1}%
{type:212-code:1-total}{1}%
{type:212-code:36-total}{1}%
{type:227-code:118-total}{1}%
{type:227-code:22-total}{1}%
{type:230-code:253-total}{1}%
{type:236-code:3-total}{2}%
{type:236-code:61-total}{1}%
{type:239-code:80-total}{1}%
{type:241-code:211-total}{1}%
{type:241-code:28-total}{1}%
{type:245-code:105-total}{1}%
{type:248-code:48-total}{1}%
{type:25-code:119-total}{1}%
{type:255-code:255-total}{12}%
{type:3-code:0-total}{17.94M}%
{type:3-code:1-total}{107.15M}%
{type:3-code:10-total}{23.07M}%
{type:3-code:109-total}{2}%
{type:3-code:11-total}{25}%
{type:3-code:13-total}{71.70M}%
{type:3-code:168-total}{1}%
{type:3-code:2-total}{51.04K}%
{type:3-code:211-total}{1}%
{type:3-code:214-total}{1}%
{type:3-code:3-total}{256.72M}%
{type:3-code:4-total}{26.66K}%
{type:3-code:6-total}{6}%
{type:3-code:7-total}{336}%
{type:3-code:8-total}{2}%
{type:3-code:9-total}{26.28K}%
{type:3-code:95-total}{1}%
{type:32-code:116-total}{1}%
{type:35-code:3-total}{1}%
{type:4-code:0-total}{46.18K}%
{type:46-code:48-total}{1}%
{type:5-code:0-total}{105.78K}%
{type:5-code:1-total}{18.01M}%
{type:50-code:46-total}{5}%
{type:69-code:0-total}{1}%
{type:74-code:58-total}{1}%
{type:8-code:0-total}{3.12M}%
{type:8-code:123-total}{1}%
{type:8-code:19-total}{96}%
{type:8-code:3-total}{4}%
{type:8-code:82-total}{3}%
{type:8-code:9-total}{739}%
{type:91-code:144-total}{1}%
}%
[\textcolor{red}{#1 not found in icmpFULL}]%
} 
\begin{document}
\frontmatter          %
\mainmatter              %
\title{Hidden Treasures -- Recycling Large-Scale Internet Measurements to Study the Internet's Control Plane}
\titlerunning{Hidden Treasures --- Recycling Large-Scale Internet Measurements to Study the Internet's Control Plane}  %
\author{Jan R\"uth \and Torsten Zimmermann \and Oliver Hohlfeld}

\authorrunning{R\"uth et al.} %
\tocauthor{Jan R\"uth, Torsten Zimmermann, Oliver Hohlfeld}

\institute{RWTH Aachen University \email{\{lastname\}@comsys.rwth-aachen.de} \\ \url{https://icmp.netray.io}} 

\maketitle              %

\begin{abstract}
Internet-wide scans are a common active measurement approach to study the Internet, \eg{} studying security properties or protocol adoption.
They involve probing large address ranges (IPv4 or parts of IPv6) for specific ports or protocols.
Besides their primary use for probing (\eg{} studying protocol adoption), we show that---at the same time---they provide valuable insights into the Internet control plane informed by ICMP responses to these probes---a currently unexplored secondary use.
We collect one week of ICMP responses (\icmp{total} messages) to several Internet-wide ZMap scans covering multiple TCP and UDP ports as well as DNS-based scans covering $>50\%$ of the domain name space.
This perspective enables us to study the Internet's control plane as a by-product of Internet measurements.
We receive ICMP messages from $\sim$171M different IPs in roughly 53K different autonomous systems.
Additionally, we uncover multiple control plane problems, \eg{} we detect a plethora of outdated and misconfigured routers and uncover the presence of large-scale persistent routing loops in IPv4.

\end{abstract}

\section{Introduction}
\label{sec:intro}
Internet scans are a valuable and thus widely used approach to understand and track the evolution of the Internet as one of the most complex systems ever created by humans.
They are widely applied in different fields, including networking and security research: \eg{} to find vulnerable systems~\cite{heartbleed14}, to measure the liveness of IP addresses~\cite{richter18}, or to measure the deployability of new protocols, features~\cite{copycat}, or their evolution~\cite{varvelloH2}.
Advances in scanning methodologies enabled probing the entire IPv4 address space for a single port within minutes or hours, depending on the available bandwidth and configured scan rate (see tools such as ZMap~\cite{Durumeric13} or MASSCAN~\cite{masscan}).
Thereby, regular scans of the entire IPv4 address space have become feasible, \eg{} providing an insightful perspective into protocol evolution (see \eg{} QUIC~\cite{ruethPAMquic}).
This line of scan-based works has created a rich body of contributions with valuable insights into Internet structure and evolution.
These works have in common that they focus on one particular feature or protocol as their objective to study ({\em primary use}).

In this work, we argue that Internet-wide scans have a less explored {\em secondary use} to study the Internet control plane while scanning for their primary use, \eg{} to detect routing loops while {\em primarily} probing for QUIC-capable servers.
That is, we study Internet control plane responses sent via ICMP as response to non-ICMP probe packets (\eg{} QUIC) and show that Internet-wide scans are a hidden treasure in that they produce a rich ICMP dataset that is currently neglected, \eg{} to uncover network problems.
The interesting aspect is that these ICMP-responses are a valuable secondary use that is generated as by-product of any Internet-wide scan.
They thus enable to study the Internet control plane (\eg{} to detect routing loops) without requiring dedicated scans (as performed a decade ago~\cite{diotloops, loopstudy}) that would increase the scanning footprint.

Our observations on the Internet's control plane are fueled by regular ZMap scans of the IPv4 address space for multiple TCP and UDP ports as well as DNS-based scans of top lists and zone files for mainly TLS, HTTP/2, and QUIC.
We evaluate one full week of ICMP responses to multi-protocol Internet-scans covering the entire IPv4 address space and $>50\%$ of the domain name space(base domains). %

Our contributions are as follows:
\begin{itemize}[noitemsep,topsep=5pt,leftmargin=10pt]
\item We propose to use Internet-wide scans to study the Internet control plane via ICMP response, \eg{} to detect routing loops or misconfigurations.
\item Within our one week observation period, we collect $\sim$\icmp{total} ICMP messages which we make available at~\cite{dataset}.
\item We shed light on how Internet-scans trigger ICMP responses across the Internet.
\item Our data shows a plethora of misconfigured systems \eg{} sending ICMP redirects across the Internet or producing deprecated source quench messages.
\item We find many networks and hosts to be unreachable, our scans uncover large sets of unreachable address space due to routing loops.
\item We provide a growing ICMP dataset at \url{https://icmp.netray.io}.
\end{itemize}
\afblock{Structure.}
The next section (\sref{sec:overview}) starts by providing an overview of our ICMP dataset.
Following this, we dive into our dataset and dissect it (\sref{sec:content}).
Driven by our findings, we inspect unreachable hosts due to routing loops and quantifies their presence in today's Internet (\sref{sec:loops}).
Finally, we discuss related works (\sref{sec:rw}) and conclude the paper (\sref{sec:conclusion}).

\section{Scan Infrastructure \& Dataset}
\label{sec:overview}

\begin{table}[t]
\newcolumntype{C}[1]{>{\centering\arraybackslash}p{#1}}
\centering
\footnotesize
\begin{tabular}{@{}C{1.5cm}||C{1.4cm}|C{1.4cm}|C{1.4cm}|C{1.4cm}|C{1.4cm}|C{1.4cm}|C{1.4cm}@{}}
\toprule
\multicolumn{1}{r||}{} & Mon & Tue & Wed & Thu & Fri & Sat & Sun \\ \midrule
\multicolumn{1}{l||}{Source} & \multicolumn{7}{c}{DNS} \\ \midrule
\begin{tabular}[l]{@{}l@{}}Protocols \& \\ Ports\end{tabular} & \multicolumn{7}{c}{TCP/443, gQUIC/443} \\ \cmidrule{1-8}\morecmidrules\cmidrule{1-8}
\multicolumn{1}{l||}{Source} & Alexa & 1\% IPv4 & \multicolumn{4}{c|}{IPv4} &  \\ \midrule
\begin{tabular}[l]{@{}l@{}}Protocols \&\\ Ports\end{tabular} & \begin{tabular}[c]{@{}l@{}}TCP/80,\\ TCP/443\end{tabular} & \begin{tabular}[c]{@{}l@{}}TCP/80,\\ TCP/443\end{tabular} & TCP/80 & iQUIC/443 & gQUIC/443 & TCP/443 & \\
\bottomrule
\end{tabular}
\caption{Weekly scan schedule fueling our dataset, DNS-based scans use our own resolver infrastructure. For IPv4-wide scans, we utilize ZMap.}
\label{tab:scan_schedule}
\vspace{-2em}
\end{table}

\afblock{Infrastructure.}
Our scans are sourced by two different modes, on the one hand, we use the ZMap~\cite{Durumeric13} port scanner on multiple machines to perform different scans over the course of a week, and on the other hand, we continuously probe $>50\%$ of the DNS space.
Table~\ref{tab:scan_schedule} summarized our weekly scan schedule, we did not specifically create these scans and this schedule for this paper, it is the result of ongoing research efforts.

These scans typically involve scanning TCP/80 for TCP initial window configurations~\cite{ruethIWIMC} or TCP fast open support.
Further, we investigate TCP/443 for HTTP/2-support~\cite{h2Push} and TLS, additionally, we scan on UDP 443 for Google QUIC (gQUIC)~\cite{ruethPAMquic} and IETF-QUIC (iQUIC).
Our DNS-based scans are fueled by using our own resolvers to resolve various record types for domains listed in zone files of multiple TLDs (\eg{} .com, .net, .org), which can be obtained from the registries, and we use A-records to investigate TLS, HTTP/2, and gQUIC.
All of our scans including the DNS resolutions originate from a dedicated subnet.
To collect all ICMP traffic that is directed towards these hosts, we install a mirror port at their uplink switch and filter it to only contain ICMP traffic that belongs to our measurement network.
Since we perform no measurements that generate ICMP messages themselves, we exclude those sent from our host (only ping responses) leaving us with only incoming ICMP traffic.

\begin{figure}[t]
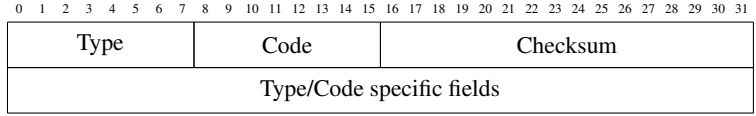

\centering
\begin{bytefield}{32}
\bitheader{0-31} \\
\bitbox{8}{Type} & \bitbox{8}{Code} & \bitbox{16}{Checksum}\\
\wordbox{1}{Type/Code specific fields}
\end{bytefield}
\caption{ICMP header structure. Type and this type's sub type (code) determine message contents, \eg{} often packets triggering the ICMP message are quoted.}
\label{fig:icmp_header}
\vspace{-1em}
\end{figure}

\afblock{Dataset.}
We base our observations on one full week in September 2018.
In this week we received \unit[169]{GB} resp. $\sim$\icmp{total} ICMPv4\footnote{Please note that we do not have a fully IPv6-capable measurement infrastructure and thus focus on IPv4 only.} messages (excluding those explicitly triggered in \sref{sec:loops}).
ICMP messages follow the structure shown in \fig{fig:icmp_header}, they are fundamentally made up of a type field and, to further specify a subtype, a code field, and depending on their value additional information may follow.

\begin{table}[]
\newcolumntype{L}[1]{>{\raggedright\arraybackslash}p{#1}}
\newcolumntype{C}[1]{>{\centering\arraybackslash}p{#1}}
\newcolumntype{R}[1]{>{\raggedleft\arraybackslash}p{#1}}
\begin{tabular}[t]{@{}l|r|r|r@{}}
\toprule
Type & Count & Uniq. IP & Uniq. AS\\ \midrule
Dest. Unreach. & \icmp{type:3-total} & \icmpNetDetails{type:3-UIPs} & \icmpNetDetails{type:3-UASs} \\
\hline
TimeExceeded & \icmp{type:11-total} & \icmpNetDetails{type:11-UIPs} & \icmpNetDetails{type:11-UASs} \\
\hline
Redirect & \icmp{type:5-total} & \icmpNetDetails{type:5-UIPs} & \icmpNetDetails{type:5-UASs} \\
\hline
EchoRequest & \icmp{type:8-total} & \icmpNetDetails{type:8-UIPs} & \icmpNetDetails{type:8-UASs} \\
\hline
SourceQuench & \icmp{type:4-total} & \icmpNetDetails{type:4-UIPs} & \icmpNetDetails{type:4-UASs} \\
\bottomrule
\end{tabular}
\hfill
\begin{tabular}[t]{@{}l|r|r|r@{}}
\toprule
Type & Count & Uniq. IP & Uniq. AS\\ \midrule
EchoReply & \icmp{type:0-total} & \icmpNetDetails{type:0-UIPs} & \icmpNetDetails{type:0-UASs} \\
\hline
Other & \icmp{type:Other-total} & \icmpNetDetails{type:Other-UIPs} & \icmpNetDetails{type:Other-UASs} \\
\hline
TimestampReq. & \icmp{type:13-total} & \icmpNetDetails{type:13-UIPs} & \icmpNetDetails{type:13-UASs} \\
\hline
Param.Problem & \icmp{type:12-total} & \icmpNetDetails{type:12-UIPs} & \icmpNetDetails{type:12-UASs} \\
\hline
Addr.MaskReq. & \icmp{type:17-total} & \icmpNetDetails{type:17-UIPs} & \icmpNetDetails{type:17-UASs} \\
\bottomrule
\end{tabular}
\hfill
\\ \caption{ICMP types with their occurrence frequency in our dataset. Ordered by frequency.}
\label{tab:overview}
\vspace{-3em}
\end{table}

\section{Study of ICMP Responses}
\label{sec:content}
To begin our investigations, we first summarize the ICMP responses to our scans by looking at the distribution of ICMP message types and their frequency of occurrence in \tab{tab:overview}.
We observe \icmp{num} different ICMP type/code combinations during our observation period with significantly different occurrence frequencies.
While we mostly receive standardized ICMP messages, we also receive some messages for which we could not find a standard, summarized as \emph{Other} in \tab{tab:overview}, on which we do not further focus in this paper.
The table lists the total count of these messages as well as the number of unique source IPs (router/end-host IPs) that generated the messages and number of ASes they are contained in.
Over the course of the week, we run different scans.
Notably, on Sundays and Mondays (see Table~\ref{tab:scan_schedule}), no IPv4-wide ZMap scans are performed.

\begin{figure}
\vspace{-1.5em}
\includegraphics{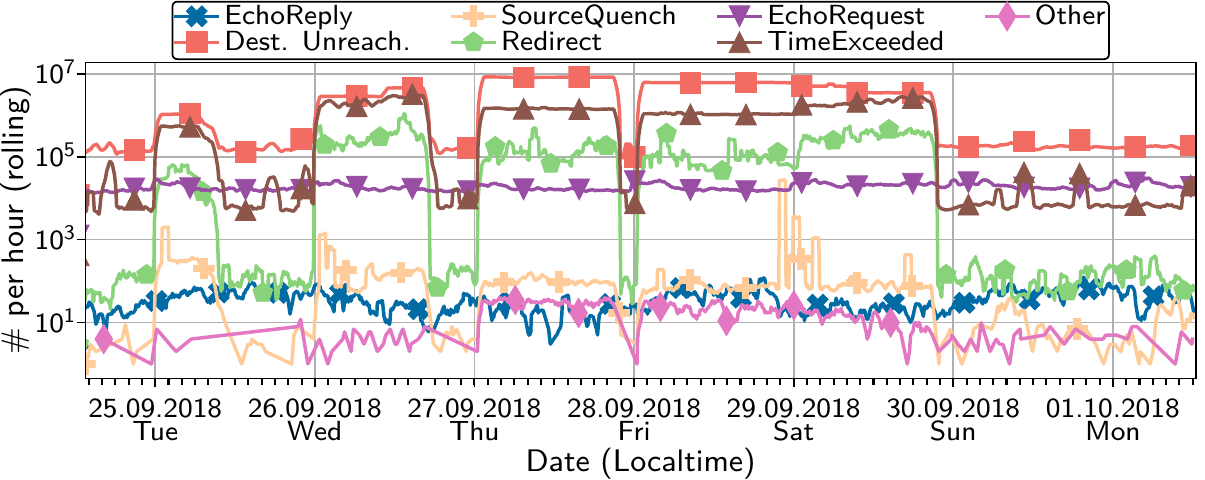}
\caption{Number of ICMP messages receiver per hour and type over the course of a week. Note the log scale and that we used a rolling sum over 1h.}
\label{fig:overview}
\vspace{-1.5em}
\end{figure}

\fig{fig:overview} thus puts the data from \tab{tab:overview} into a temporal context showing the rolling sum over 1h intervals of the major ICMP types.
We observe that the ICMP traffic varies over the course of the week, \eg{} echo requests are rather static, other types like destination unreachable mainly follow our ZMap scan schedule.

\afblock{Quoted IP Packet.}
Apart from the different ICMP types, many ICMP messages contain parts of the packet that caused the creation of the messages.
We further inspect these quoted IPv4 packets within the ICMP messages.
From all received ICMP messages, 99.5\% are supposed to contain IP packets (according to the RFCs), of these only 0.07\% cannot be decoded, \eg{} because there is simply not enough data or these are no IPv4 packets.
Of the decodable packets, we find 180.25M unique source IP/payload length combinations, 76\% are longer than 40 bytes, \ie{} enough to inspect IP and TCP headers when no options are used\footnote{To reduce the capture size, our packet capture caps packets at 98 byte allowing no further investigation, we find 67\% having the maximum capture size.}, 24\% are exactly 28 byte long, so just enough to inspect the transport ports.
Thus, when no options are used, the chances are high that ICMP messages received by an ICMP receiver can be demultiplexed to the respective application process.
This extends the finding in~\cite{maloneICMPQuotation} that showed a prevalence of 28 byte responses for TCP \texttt{traceroutes}.
Next, we focus on the destination address field within the quoted IP header.
These should correspond to addresses which are targeted by our scanners.
Interestingly, from all ICMP messages, we find over 1.06M messages with destination IPs that are in reserved address space, \ie{} unallocated or private addresses (\eg{} 192.168.0.0/24).
Since all our scanners explicitly blacklist these IP addresses, we believe that these messages are produced behind network address translations (NATs).
We next use the contained source addresses to understand the relation to our measurements.

\takeaway{ICMP traffic shows a temporal correlation to measurement traffic, most messages indicate unreachability. 
In our collected dataset, quoted IP packets typically contain enough information to inspect everything up to the end of the TCP header.
Further, a substantial number of messages seems to be generated behind NATs allowing to peek into private address spaces.}

\subsection{Responses to Individual Measurements}
\label{sec:measurementimpact}
Since we perform a variety of different measurements independent of this study, our first investigation is how different measurements affect the generation of ICMP traffic.
To this end, we compare two ZMap scans and a purely DNS-based scan.
For the ZMap scans, we focus on one that enumerates reachable TCP port 80 (HTTP) and UDP port 443 (QUIC) hosts, for DNS, we use a scan that probes for HTTP/2 support via TCP port 443.
We are able to clearly tie the ICMP messages to the different scans via IPs and ports either from the quoted IP message or from IP itself.

\begin{figure}[t]
\includegraphics{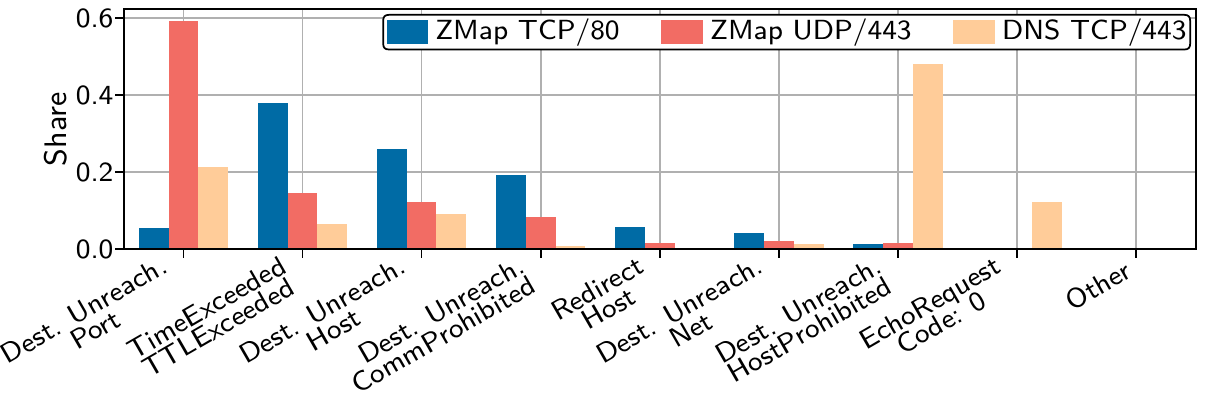}
\vspace{-2em}
\caption{ICMP messages triggered by ZMap and DNS-based scans.}
\label{fig:zmapdns_cmp}
\vspace{-1.5em}
\end{figure}

\fig{fig:zmapdns_cmp} shows the distribution of ICMP types \emph{and} codes (top 8) that we receive for the respective scans.
As already indicated by \tab{tab:overview}, we receive a large amount of destination unreachable messages. 
However, depending on the scan, their amount and share greatly vary, especially when looking at the respective code.
For example, unreachable ports are very common for our UDP-based ZMap scan, however, in comparison, the TCP-based ZMap scan shows only a small fraction of unreachable ports.
This is no surprise as TCP should reply with a RST-packet if a port is unreachable and does typically not generate ICMP messages.
In contrast, there is no such mechanism in UDP, even through something comparable to TCP's RST exists in QUIC.
However, QUIC is implemented in user-space, thus when the kernel cannot demultiplex a packet to a socket it must resort to issuing an ICMP unreachable message.
Looking at our DNS-based scan, we still find that more than 20\% of the ICMP messages signal unreachability through ICMP in contrast to TCP RSTs, something that, \eg{} the default ZMap TCP-SYN scan module simply ignores in contrast to its UDP counterpart.
Since in all major operating systems TCP handles signaling closed ports, we believe that these hosts issuing ICMP replies are actively configured either in their own firewalls (\eg{} iptables) or in a dedicated firewall to do so.
We find only \dnshtwodetail{type:3-code:3-UIPs} IPs issuing \emph{all} \dnshtwo{type:3-code:3-total} ICMP port unreachable messages, supporting our assumption that dedicated machines filter this traffic.

Looking at the other types/codes, we find that a non-negligible share of ICMP messages indicate that hosts are not reachable via the Internet either due to TTLs expiring or because their host or network cannot be reached.
Apart from this, we observe that TCP port 443 is often firewalled (HostProhibited).

\takeaway{Depending on the protocol and port, we get different feedback from the Internet's control plane. Our findings indicate that, \eg{} ICMP port unreachable messages should not be ignored for TCP-based scans as is currently the case.}

\subsection{ICMP Echos}
ICMP echo requests (Type: 8) are the typical \texttt{ping} to which an echo reply is sent.
RFC792 defines only a single code point, \ie{} code = 0 which represents ``no code'', still we observe some non-standard code points.
Some security scanners use non-standard code points for operating system fingerprinting, \eg{} a standard Linux will echo the requested code point in its reply.
Still, pings to our measurement infrastructure seem quite common, for code = 0, we find \icmpNetDetails{type:8-code:0-UIPs} unique IPs out of \icmpNetDetails{type:8-code:0-UASs} autonomous systems (ASes).
It seems that our scanning activities trigger systems to perform ping measurements towards us, yet, we do not know their actual purpose.
We suspect that this could be caused by intrusion detection systems (IDSs) that monitor the liveness of our hosts.

\afblock{Echo Replies.}
Since our hosts do not perform echo requests, we were surprised to find echo replies in our dataset.
We observe different code points with different frequencies but overall we find over a couple of thousand of these replies.
To investigate what causes these seemingly orphaned messages, we inspect their destinations.
Since our measurements are identifiable either by IP or additionally by weekday, we associate messages to measurements.
We find most echo replies are with code = 3 (except for 5 messages), all \icmp{type:0-code:3-total} of these echos are destined to our DNS resolvers and originate from only 86 IP addresses in 2 Chinese ASes.
While many ICMP packets contain IP quotations, echo replies typically do not, they usually mirror data contained in the echo request.
Yet, we still find IP packets together with DNS query \emph{responses} that are destined to our resolver.
Thus, it seems that the packets are generated on the reverse path, however, they are not sent back to the source (DNS server) but they are forwarded to the destination (us).
Inspecting the source IP within the IP fragments, we find IP addresses from the same two ASes, as it turns out the \icmpNetDetails{type:0-code:3-UIPs} ICMP source IPs all respond to DNS queries which hints at their use as a DNS server cluster.
Yet, we were unable to manually trigger these ICMP reply packets when trying to send DNS requests to these IPs, we only observed that DNS requests were always answered by two separate packets from the same IP, however, with different DNS answers.
Further, the packets seem to stem from different IP stacks (significantly different TTLs, use of IP ID or not, don't fragment bit set or not).
While the different stack fingerprints could be the result of middleboxes altering the IP headers, the general pattern that we observe hints at DNS spoofing.

\subsection{Source Quench}
ICMP Source Quench (SQ) messages (Type: 4, Code: 0) were a precursor of today's ECN mechanism, used to signal congestion at end-hosts and routers.
The original idea (RFC792~\cite{RFC792}) was that a router should signal congestion by sending SQ messages to the sources that cause the congestion.
In turn, these hosts should react, \eg{} by reducing their packet rate.
However, research~\cite{sqcc} found that SQ is ineffective in \eg{} establishing fairness and IETF has deprecated SQ-generation in 1995~\cite{RFC1812} and SQ-processing in 2012 in general~\cite{RFC6633}.
Major operating systems ignore SQ-messages for TCP at least since 2005 to counter blind throughput-reduction attacks~\cite{RFC5927}.
Further,~\cite{floydECN} claims that SQ is rarely used because it consumes bandwidth in times of congestion.

In our traces, we observe \icmpNetDetails{type:4-code:0-UIPs} unique IPs located in \icmpNetDetails{type:4-code:0-UASs} ASes issuing SQ messages, despite the deprecation.
Out of these IPs, \unit[34.42]{\%} are located in only 5 ASes.
Moreover, 609 IPs that generate SQ messages were directly contacted by our measurement infrastructure, \ie{} are the original destination of the request causing this SQ message (according to the IPv4 header contained within the ICMP message).
Among the remaining SQ messages, we find a few messages where the original destination and the source of the SQ messages are located in ASes of different operators, \ie{} possible transit networks.
Exemplarily, we observe that IPs located in AS1668 (AOL Transit Data Network) and AS7018 (AT\&T) issued SQ messages when IPs located in AS8452 (Telecom Egypt) were contacted.
As a final step, we see that 53 destination IPs in our measurements trigger the generation of SQ messages and are also contained in A-records of our DNS data that we collect.
Out of these 53 IPs, 22 IPs generated the SQ messages themselves, \ie{} no on-path intermediary caused the creation of this message.

In addition, we checked how vendors implement or handle this feature.
Cisco removed the SQ feature from their IOS system after Version 12 in the early 2000s~\cite{ciscoquench}.
Hewlett Packard's cluster management system (Serviceguard) generated SQ messages due to a software bug in a read queue, which was fixed by a patch in 2010~\cite{hpserviceguard}.
In their router configuration manual (September 2017), Nokia also marks SQ messages as deprecated~\cite{nokiamanual}.
Although we cannot identify devices and their operating system version in our measurements, we assume that some devices are not updated to a current version or are following a configuration that enables them to generate SQ messages. 
This is not forbidden per se but given that ICMP SQ creation was deprecated over 20 years ago, our findings highlight that removing features from the Internet is a long term endeavor.

\subsection{Redirect}
ICMP redirect messages (Type: 5), are sent by gateways/routers to signal routes to hosts.
While~\cite{redirectbaad} finds networks which require redirect messages to be architected sub-optimally in the first place, RFC1812~\cite{RFC1812} states that a router \emph{must not} generate redirect messages unless three properties are fulfilled:
\emph{i)} The packet is being forwarded out the same physical interface that it was received from, \emph{ii)}, the IP source address in the packet is on the same logical IP (sub)network as the next-hop IP address, and \emph{iii)}, the packet does not contain an IP source route option.
Similar checks~\cite{RFC1122} are used by receiving hosts to check the validity of the message (\eg{} redirected gateway and issuing router must be on the same network).

Since none of the \icmp{type:5-total} redirect messages originate from our network, the routers generating them either violate rule \emph{ii)} or some obscure address translation is in place on their networks.
In our data, we even find roughly 2.7K unique redirects to private address space.
Within our dataset, we observed \icmp{type:5-code:0-total} network redirects and \icmp{type:5-code:1-total} host redirects.
Network redirects are problematic since no netmask is specified and it is up to the receiving router to interpret this correctly.
For this reason, RFC1812~\cite{RFC1812} demands that routers \emph{must not} send this type.
We find that the network redirects originate from \icmpNetDetails{type:5-code:0-UASs} different ASes affecting nearly 19k different destinations of which less than 20 are mapped in any of our DNS data.
Yet, all these ASes thus contain questionable router configurations that are outdated at least since 1995.
Similarly, we find that the much larger fraction of host redirects originate from \icmpNetDetails{type:5-code:1-UASs} ASes that affected over 400k destinations of which we find roughly 900 mapped in our DNS data.
This suggests that a substantial number of end-systems are connected via sub-optimally architected or misconfigured networks.

\begin{table}[t]
\newcolumntype{L}[1]{>{\raggedright\arraybackslash}p{#1}}
\newcolumntype{C}[1]{>{\centering\arraybackslash}p{#1}}
\newcolumntype{R}[1]{>{\raggedleft\arraybackslash}p{#1}}
\begin{tabular}[t]{@{}L{2.0cm}|L{2.3cm}|R{1.3cm}@{}}
\toprule
Type & Code & Count\\ \midrule
Dest. Unreach. & Port & \icmp{type:3-code:3-total} \\
\hline
TimeExceeded & TTLExceeded & \icmp{type:11-code:0-total} \\
\hline
\multirow{5}{*}{Dest. Unreach.} & Host & \icmp{type:3-code:1-total} \\
\cline{2-3}
 & CommProhibited & \icmp{type:3-code:13-total} \\
\cline{2-3}
 & HostProhibited & \icmp{type:3-code:10-total} \\
\cline{2-3}
 & Net & \icmp{type:3-code:0-total} \\
\cline{2-3}
 & Protocol & \icmp{type:3-code:2-total} \\
\bottomrule
\end{tabular}
\hfill
\begin{tabular}[t]{@{}L{2.0cm}|L{2.3cm}|R{1.3cm}@{}}
\toprule
Type & Code & Count\\ \midrule
\multirow{2}{*}{Dest. Unreach.} & Frag.Needed & \icmp{type:3-code:4-total} \\
\cline{2-3}
 & NetProhibited & \icmp{type:3-code:9-total} \\
\hline
TimeExceeded & Frag.Reassembly & \icmp{type:11-code:1-total} \\
\hline
\multirow{4}{*}{Dest. Unreach.} & HostUnknown & \icmp{type:3-code:7-total} \\
\cline{2-3}
 & NetTOS & \icmp{type:3-code:11-total} \\
\cline{2-3}
 & NetUnknown & \icmp{type:3-code:6-total} \\
\cline{2-3}
 & SourceIsolated & \icmp{type:3-code:8-total} \\
\bottomrule
\end{tabular}
\hfill
\\
 \caption{ICMP messages received indicating some form of unreachability with known type and code ordered by frequency.}
\label{tab:unreach}
\vspace{-3em}
\end{table}

\subsection{Unreachable Hosts}
Reachability is a fundamental requirement to establish any means of communication.
Given that \tab{tab:overview} lists \icmp{type:3-total} destination unreachable messages this looks troublesome at first.
Yet, not all unreachability is bad, \eg{} firewalls actively protect infrastructure from unpermitted access, \ie{} when iptables \emph{rejects} a packet (in contrast to simply dropping it) it generates an ICMP response.
By default, a port unreachable message (Type: 3, Code: 3) is produced but other types can be manually specified by the network operator.
Our scans in themselves certainly trigger a certain amount of firewalls or some IDSs.
In contrast, when a path is too long and the IP TTL reaches zero, routers typically generate an ICMP TTL exceeded message indicating that the destination is not reachable but this time due to the network's structure.
Similarly, ICMP destination unreachable messages for host unreachable (Type: 3, Code: 1) should indicate that there is currently simply no path to a host, \eg{} because it is not connected or the link is down.
\tab{tab:unreach} summarizes the unreachability that we observe in our dataset.

As already indicated in \sref{sec:measurementimpact}, our UDP-based ZMap scans have the highest share of port unreachable messages putting them at the top.
We inspect the origin of the messages and the actual destination that our scans targeted to see if the end-hosts generate the messages or an intermediate firewall.
It seems that 96\% of the messages are indeed generated by end-hosts or machines that can answer on their behalf (NATs).

\afblock{Host and Network.}
Unreachable hosts and networks codes are used to give hints that currently no path is available and the RFCs explicitly note that this may be due to a transient state and that such a message is not proof of unreachability.
To check for transient states, we compare the unreachable hosts on Thursday with those on Friday in our ZMap (both UDP 443) scan and additionally with the same scan (Thursday) one week later (captured separately from our initial dataset) and investigate if hosts become reachable that were unreachable before or vice versa.

\begin{figure}[t]
\begin{subfigure}{0.5\textwidth}
\includegraphics{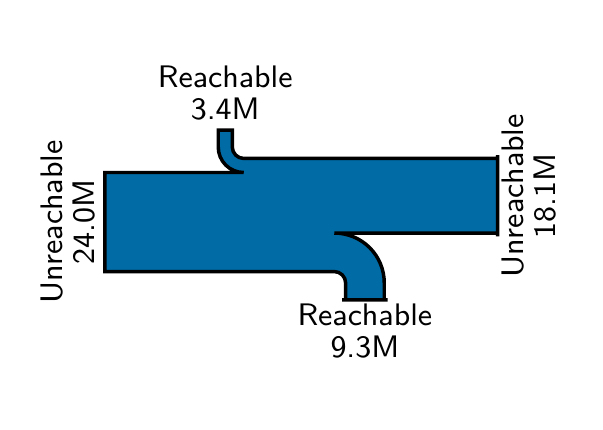}
\vspace{-4em}
\caption{Thursday to Friday.}
\end{subfigure}
\begin{subfigure}{0.5\textwidth}
\includegraphics{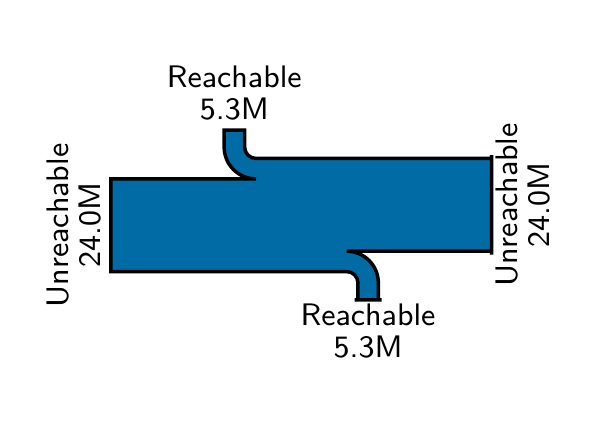}
\vspace{-4em}
\caption{Thursday to Thursday one week later.}
\end{subfigure}
\vspace{-0.5em}
\caption{Different scans (left to right of each plot) trigger different amount of host unreachable messages. (a) compares the changes within one day. (b) within one week.}
\label{fig:reachablesankey}
\vspace{-1.5em}
\end{figure}

\fig{fig:reachablesankey} visualizes the change between these two days (a) and within one week (b) for host unreachable messages.
We can see that within two days, the majority of hosts remain unreachable, a small number of hosts that were previously \textit{reachable}\footnote{With reachable we actually mean \textit{not unreachable}, \ie{} we do not get ICMP unreachable messages, which must not mean that this host was reached by the scan.} become unreachable, similarly, previously unreachable hosts become reachable.
Looking at the changes within a full week, we observe that the total amount of unreachable hosts stays the same, however, roughly the same amount of previously reachable host become unreachable and vice versa.
To dig into these once unreachable and then reachable hosts, we inspect to which AS they belong finding that 82\% of all hosts are from the same ASes.
A possible explanation might be that while our observations seem to indicate a change, the ICMP message generation is subject to rate-limiting~\cite{guoPAM18}.
Thus there might be routers that generated unreachable messages on Thursday for a certain host, however, this router could be subject to rate-limiting on Friday for the same host or the week after leading to a false impression of reachability and continuity, still, a substantial number of hosts remain unreachable.
Another possibility is that some hosts are only up at certain times of the day leading to differences in the reachability.

\afblock{Time Exceeded.}
Similar to host unreachability, Time Exceeded messages (Type:11) indicate unreachability but due to network issues.
Either the Fragment Reassembly (Code: 1) time was exceeded, \ie{} the time that IP datagrams are buffered until they can be reassembled when IP fragmentation happens, or the TTL runs out (Code: 0), \ie{} the path length exceeds the sender-defined limit.
For the former, we find some thousand messages but they stem from only \icmpNetDetails{type:11-code:1-UASs} ASes, since many of our scans use small packets, fragmentation is unlikely in the first place.
Yet, for example, the UDP ZMap scans use roughly 1300 byte per packet which is in the range of typical~\cite{anaTMA2018} MTUs when fragmentation could occur.
Since the default ZMap functions to create IP packets (which we use) do not set the \textit{don't fragment bit}, only some of our measurements trigger the \icmp{type:3-code:4-total} \textit{fragmentation needed and DF set} ICMP messages (see \tab{tab:unreach}).
However, over time, these ICMP messages could give valuable insights into path MTU in the Internet.

TTL Exceeded messages have the second largest occurrence (\icmp{type:11-code:0-total}) within our dataset.
They were produced in \icmpNetDetails{type:11-code:0-UASs} different ASes covering 35.5M different destinations that our scans tried to reach of which $\sim$32K are again present in A-records of our DNS data and are thus unreachable.
We inspect the TTL field of the quoted IP packets that triggered the ICMP messages to see if the TTL was actually zero when the message was generated.
To do so, we first generate all unique pairs of router IP and TTL values and then count the different TTLs observed.
Out of these, 97\% of the TTLs show a value of one, followed by $\sim$2.4\% with a zero, we expect these two, since a router should drop a TTL = 0 or, depending on the internal pipeline, also TTL = 1, when the packet is to be forwarded.
Nevertheless, we also find larger TTLs, 2, 3, 4, 5, and 6 directly follow in frequency, yet, we also find some instances of over 200 or even 255.
The very large TTLs could hint at middleboxes or routers rewriting the TTL when they generate the message to hide their actual hop count.
The lower numbers could be indicators for MPLS networks.
By default, \eg{} Cisco~\cite{ciscoMPLSTTLPropagation} and Juniper~\cite{juniperMPLSTTLPropagation} routers copy the IP TTL to the MPLS TTL on ingress and also decrement the IP TTL within the MPLS network.
It is possible to separate IP TTL and MPLS TTL and there are heated discussions whether one should hide the MPLS network from traceroutes or not which has also been subject of investigations~\cite{luckieMPLStunnels}.
Thus packets expiring within an MPLS network will still trigger an ICMP TTL exceeded, however, the quoted IP packet will have the TTL value they had at the MPLS ingress router, thus, if the IP TTL is still copied at ingress a traceroute could still reason about an MPLS network.

Since we were surprised to see this many TTL exceeded messages across all scanner types (see \sref{sec:measurementimpact}), we checked our scanners to see which TTL they were actually using to see if our setup simply has too small values.
All our ZMap-based scanners initialize the TTL field with its maximum of 255 possible hops, all scanners building on top of the transport layer interfaces, in contrast, use the current Linux default of 64 hops as also recommended in RFC1700~\cite{RFC1700}.
Given that we are at least on the recommended hop count, this leaves us with three possibilities, \emph{i)} the current recommendation of 64 is too low to reach these hosts, \emph{ii)} there are middleboxes modifying the TTL to a much lower value, or, \emph{iii)} there are routing loops on the path to these hosts.
After shortly summarizing our findings, we continue by exploring the latter.

\subsection{Summary}
As the previous sections have shown, our Internet-wide scans produce an insightful \emph{secondary} dataset of ICMP responses.
Driven by these messages, we identified a potential DNS spoofer, found that long deprecated source quench messages are still generated in today's Internet and that ICMP redirects are sent across different administrative domains pointing to several outdated and misconfigured networks.
Without crafting a dedicated dataset, our scans enable us to study Internet reachability and we believe that longitudinal studies offer a way to deal with the challenge of ICMP rate-limiting.

\section{Routing Loops}
\label{sec:loops}
Routing loops are an undesirable control plane misconfiguration that render destination networks unreachable and that challenge a link's load~\cite{floodingloops}.
In essence, IP's TTL protects the Internet from indefinitely looping packets and thus ICMP TTL messages inform the sender that a router dropped a packet after exceeding the allowed number of router hops (TTL).
While the potential for routing loops is known, only a few studies investigated their presence a decade ago~\cite{diotloops, loopstudy}, current information on the presence and prevalence is missing.
Therefore, we study routing loops on the basis of ICMP TTL exceeded messages triggered by our scans.
We further argue that routing loops can be frequently investigated as a by-product of Internet-wide scans that are regularly conducted for different purposes.

\subsection{Methodology: Detecting Loops}
\label{sec:method}
ICMP TTL exceeded messages are not necessarily caused by loops, also overly large paths or middleboxes could trigger these messages.
To investigate whether or not an actual loop is present, we perform \texttt{traceroutes} for the original destinations (in the quoted IP) of the ICMP TTL exceeded messages.
Since our traceroutes are subject to ICMP rate-limiting, especially when packets start to loop, we customize traceroute.
Our traceroute slows down its sending rate when detecting an already seen IP address (loop indicator), otherwise, it follows the design of \textit{Paris traceroute}~\cite{paristraceroute} reusing flow identifiers for each hop to trigger the same forwarding behavior in ECMP-like load balancers.

Since the traceroutes can still be noisy due to hosts that do not generate ICMP at all or are still subject to rate-limiting, especially when also other traffic flows into a loop, we put strong demands on our loop.
For each hop on the path that does not generate a reply, we assign a new unique label, all others are simply labeled by the answering IP.
From this list of labels, we create a directed graph connecting each label-induced node to its successor and on this path we compute all elementary cycles using~\cite{elemCycles}.
On an elementary cycle, no node appears twice except that the first and last node are the same.
Then, on each of these possible cycles we inspect the node with the highest degree, and if this node's degree is larger than 5\footnote{This is basically a precaution against bad load balancers traded against the required TTL.}, we mark this traceroute as having a loop.
This will yield loops as long as at least one router in the loop generated ICMP TTL exceeded messages, which we found to work reasonably well when traceroute pauses the packet generation for at least \unit[500]{ms} when observing an already seen IP address.
Thus in a loop of two routers, we will send each router a packet roughly every second.

\subsection{Routing Loops in the Wild}
\label{sec:loopswild}
We seed our traceroutes by ICMP TTL exceeded messages generated from our Internet-wide scans\footnote{Our dataset excludes TTL exceeded messages generated by these traceroutes.}.
Since we get way too many TTL exceeded messages to traceroute them all without generating substantial rate-limiting, we restrict us to a single traceroute for each unique /24 subnet within 30-minute intervals.
Thus for two TTL exceeded messages for a destination from the same /24 subnet, we only perform a single traceroute if the messages arrive within 30 minutes.

For our assessment of routing loops, we investigate TTL exceeded messages in the last week of August 2018.
To avoid rate-limiting we also limit our traceroutes that we perform in parallel; generating all traceroutes for this single week took us until the end of September 2018.
While this skews our data, it enables us to reason about the persistence of these loops since every 30 minutes the same /24 could be scheduled for a rescan.
In total, we performed $\sim$27M traceroutes to $\sim$612K different /24 subnets from 28K ASes, of these, 439K subnets from 19.8K ASes are unreachable due to a loop.
We further inspect how many loops are present and if loops are only within a single AS or whether loops cross AS borders and are thus potentially on a peering link.
To do so, we count the number of distinct loops and ASes involved in the loops and find 167K different loops in 13.9K ASes.
Of these loops, 136K have IPs for all routers involved in the loop, thus allowing an in-depth inspection.
Looking at the ASes involved, we find that 13\% (17.7K) already cover all different ASes paths involved (\ie{} we replaced each IP by the respective AS), of these 4.8K cross AS boundaries.
The top three ASes involved in the loops are AS171 (Cogent) a Tier-1, AS9498 (BHARTI Airtel Ltd.), an Indian ISP, and AS3549 (Level 3), again a Tier-1.

\afblock{Persistence.}
To investigate the persistence, we restrict our view to traceroutes that were performed two weeks after the initial TTL exceeded message was triggered by our Internet-wide scans.
In contrast to our previous observation, loops from roughly 150 ASes disappear, yet, we still find 4.6K loops crossing AS borders, in total still rendering 404K subnets unreachable.
Thus, most loops seem to persist and are not resolved.

\afblock{Loops at our Upstream ISP.}
Within our data, we also found loops in the AS of our upstream ISP.
We contacted the ISP about our findings which they were able to confirm.
Since many of the loops are outside of their administrative domain even though they manage the address space, they were still able to give us more details on a loop that they were able to fix.
For one loop, they found that the first router had a static route for our tested destination towards its next hop, yet, the next hop had no specific forwarding information for this destination and thus used its default gateway, which however was the previous router with the static route thus causing the loop.

\takeaway{Routing loops seem to persist in large parts of the Internet, challenging the question if the address space cut off by the loops is in use after all or if there are other routers that would be taken from different vantage points.
We believe routing loops have a huge potential for causing congestion when exploited and thus a persistent monitoring seeded by large-scale Internet measurements that informs operators could be a long-term attempt to reduce routing loops.}

\section{Related Work}
\label{sec:rw}

Our work relates to approaches analyzing ICMP traffic and its generation in general, as well as approaches that focus on particular studies built upon ICMP, \eg{} path/topology discovery and routing loops.
In the following, we discuss similarities and differences to our work but we remark that the body of works building on top of ICMP is far larger but conceptually differ in that they do not analyze ICMP as a by-product.

Bano et al.~\cite{richter18} also use ZMap and capture \emph{all} (cross-layer) responses to probe traffic to infer IP liveness but run specific measurements to generate this traffic, we believe that our dataset could be used to perform a similar analysis.
Malone et al.~\cite{maloneICMPQuotation} analyze the correctness of ICMP quotations.
They base their analysis on a dataset obtained via \texttt{tcptraceroute} in 2005, targeting around 84K web servers. 
While most of the reported messages are of type ICMP time exceeded, they also find around 100 source quench messages, which were already deprecated then.
As we have shown, by looking at the ICMP responses to Internet-wide scans, we are able to update their findings on a regular basis without having to craft a dedicated dataset.
Guo et al.~\cite{guo2018ratelimit} present FADER, an approach to detect the presence of ICMP rate-limiting in measurement traces.
While we did not focus on rate-limiting, we found indicators for rate-limiting.
We believe that longitudinal studies seeded by Internet-wide scans can, in the long run, help to overcome limited visibility due to rate-limiting.

In 2002, Hengartner et al.~\cite{diotloops} have characterized and analyzed the presence of routing loops in a Tier-1 ISP backbone trace.
Xia et al.~\cite{floodingloops,loopstudy} have further tracerouted over 9M IP addresses to find routing loops in 2005.
Transient routing loops have also been subject to investigation~\cite{routingfailures} and they are well studied~\cite{diotdynamics,olivierloops}.
Lone et al.~\cite{lonePAMspoofloop} investigate routing loops in CAIDA data to study source address validation but do not focus on their prevalence in the Internet, further, in contrast to using the CAIDA dataset that actively runs traceroutes against all /24, we utilize indications from ongoing measurement data to investigate loops. 
While these works show that routing loops are a known problematic misconfiguration, their presence in the Internet has not been analyzed for over 10 years.
By recycling Internet-wide scans, we can seed such investigations and enable persistent monitoring of this phenomenon showing that routing loops are still a problem today.

\section{Conclusion}
\label{sec:conclusion}
In this paper, we used ICMP responses triggered by large-scale Internet measurements to study how the Internet's control plane reacts to these measurements.
Thereby, we found that these responses are hidden treasures that are typically neglected but offer great insights into the configuration of Internet-connected systems.
Our analyses of different ICMP responses led us to many misconfigured routers, \eg{} sending ICMP redirects across the Internet, or outdated systems, \eg{} generating long-deprecated source quench messages.
Further, our analysis showed a large and nuanced degree of unreachability in the Internet.
More specifically, our scans hint at the existence of routing loops, which we found to persist in large parts of the Internet.
We hope that these ICMP by-products are analyzed by more researchers when performing large-scale measurements and that the regular nature of these scans will enable persistent monitoring of the Internet's control plane and that, especially when brought to the attention of operators, misconfigurations can be fixed.
To this end, we make our dataset publicly available at~\cite{dataset}.

\section*{Acknowledgments}
Funded by the Excellence Initiative of the German federal and state governments, as well as by the German Research Foundation (DFG) as part of project  B1 within the Collaborative Research Center (CRC) 1053 –- MAKI.
We would like to thank the network operators at RWTH Aachen University, especially Jens Hektor and Bernd Kohler as well as RWTH's research data management team.

\bibliographystyle{abbrv}
\bibliography{literature.bib}

\end{document}